\documentstyle[prc,aps,amssymb]{revtex}

\begin{document}
\draft

\title{Simple model for decay of superdeformed nuclei}

\author{C.\ A.\ Stafford and B.\ R.\ Barrett}
\address{Physics Department, University of Arizona, 1118 E.\ 4th Street,
Tucson, AZ 85721}


\twocolumn[\hsize\textwidth\columnwidth\hsize\csname@twocolumnfalse\endcsname

\maketitle

\begin{abstract}
Recent theoretical investigations of the decay mechanism out of 
a superdeformed nuclear band have yielded qualitatively
different results, depending on
the relative values of the relevant decay widths.
We present a simple two-level model for the dynamics of the tunneling 
between the superdeformed and normal-deformed bands, which treats 
decay and tunneling processes on an equal footing.  The previous theoretical
results are shown to correspond to coherent and incoherent limits
of the full tunneling dynamics.  Our model accounts for experimental data
in both the $A\sim 150$ mass region, where the tunneling dynamics is
coherent, and in the $A\sim 190$ mass region, where the tunneling dynamics
is incoherent.
\end{abstract}

\pacs{PACS numbers: 
21.60.-n,
21.10.Re,
27.70.+q,
27.80.+w
}

\vskip2pc]

One of the most interesting recent discoveries in nuclear-structure physics
is the existence of superdeformation for nuclei in the mass
$A\sim 150$ and $A\sim 190$ regions.  So far, a consistent theory regarding
the decay out of a superdeformed (SD) rotational band into a normal-deformed
(ND) band has not been achieved.  Most of the early work 
\cite{schiffer,vigezzi1,vigezzi2,shimizu,shimizu2,khoo}
on this problem attributed the decay-out process to a mixing of the SD 
states with ND states of equal spin.  Decay out of the SD band sets in at a 
spin $I_0$ for which penetration through the barrier between the SD minimum
and the ND minimum is competitive with the E2 decay within the SD band.
A statistical model was used \cite{vigezzi1,vigezzi2} 
to describe the ND states, and the 
decay out of the SD band was determined as a function of the decay widths
$\Gamma_S$ and $\Gamma_N$ in the SD and ND potential wells, respectively;
the spreading (or tunneling) width $\Gamma^{\downarrow}$ 
through the barrier; and the
average spacing $D_N$ of the ND states.  Under the assumption that the ND
states form a continuum on the scale of the other energies in the problem,
the spreading width was found using Fermi's golden rule to be \cite{shimizu2}
\begin{equation}
\Gamma^{\downarrow} = 2\pi\langle V^2\rangle/D_N,
\label{gamma.wrong}
\end{equation}
where $\langle V^2\rangle$ is the mean square of the coupling matrix elements
$V_{\alpha\beta}$ connecting the SD and ND states.  $\Gamma^{\downarrow}$ 
measures the
strength of the coupling between the SD and ND states.  In Refs.\ 
\onlinecite{vigezzi1,vigezzi2}, 
it was assumed that $\Gamma_N \lesssim \Gamma_S$
and that $\Gamma^{\downarrow}/D_N \sim 1$, i.e., that 
the coupling between the SD and ND states
is relatively strong.

Quite recently, a different approach to this problem has been reported
\cite{wbb}.  In this approach, the reduction factor $F_S$ of the 
intraband transition intensity is calculated directly as a function of 
the spreading width $\Gamma^{\downarrow}$ and of the intraband E2 
width $\Gamma_S$.  Their final result for $F_S$ 
is shown to be independent of the 
statistical E1 decay widths $\Gamma_N$ of the ND states, provided that
$\Gamma_N \gg \Gamma^{\downarrow},\,\Gamma_S$.  
Since the publication of the later result,
it has been difficult to reconcile the predictions of these two calculations,
because their final results do not depend upon the same parameters.

The purpose of the present manuscript is to formulate a simple two-level
model for this problem, so as to study in detail the dependence of the
decay-out process on $\Gamma_S$, $\Gamma_N$, and $\Gamma^{\downarrow}$.  
It will be shown 
that the results of both previous investigations of the decay-out process can
be obtained in certain limits, 
depending on the relative sizes of these widths.  It will also
be shown that the appropriate expression for the spreading width is
not Eq.\ (\ref{gamma.wrong}), but
\begin{equation}
\Gamma^{\downarrow} = \frac{\displaystyle 2\bar{\Gamma} V^2}{
\displaystyle \Delta^2 +\bar{\Gamma}^2},
\label{gamma.right}
\end{equation}
where $\bar{\Gamma}=(\Gamma_S+\Gamma_N)/2$, $V$ is the matrix element
connecting the SD state of interest with the single ND state with which it
mixes most strongly, and $\Delta$ is the energy difference of 
these two states.  It will be shown that Eq.\ (\ref{gamma.wrong}) is indeed
the correct mean value of $\Gamma^\downarrow$ over a statistical ensemble of 
nuclei in the limit $V\ll \bar{\Gamma}$, in agreement with 
Ref.\ \onlinecite{wbb}.  However, the fluctuations in $\Gamma^\downarrow$ are
typically much larger than its mean, indicating that Eq.\ (\ref{gamma.right})
must be used to describe the decay out of a particular superdeformed state.

Motivated by the experimental fact \cite{kuhn}
that $\Gamma_N,\, \Gamma_S\ll D_N$ in the $A \sim 190$ region, 
we consider
an effective two-level system consisting of the superdeformed state $S$ and 
the normal-deformed state $N$ to which it couples most strongly.  
The Hamiltonian of the system is
\begin{equation}
H=H_0 + H_D,
\end{equation}
where $H_D$ describes the electromagnetic decay processes $\Gamma_N$
and $\Gamma_S$ {\em within} the
ND and SD bands, and
\begin{equation}
H_0 = \varepsilon_S c^{\dagger}_S c_S + \varepsilon_N c_N^{\dagger} c_N + 
V \left(c_S^{\dagger} c_N + c_N^{\dagger} c_S\right)
\label{h0}
\end{equation}
describes the effective two-level system, including 
tunneling through the barrier separating the SD and ND states.  Here
$c_S^{\dagger}$ and $c_N^{\dagger}$  are
creation operators for the superdeformed 
state $S$, of energy $\varepsilon_S$, and the normal-deformed state 
$N$, of energy $\varepsilon_N$, respectively.
Without loss of generality,
the tunneling matrix element $V$ is chosen positive via an appropriate choice
of the relative phase of the states $S$ and $N$.

In order to include both 
the coherent ``Rabi oscillations'' due to $V$ and the irreversible
decays $\Gamma_S$ and $\Gamma_N$, it is useful to consider the retarded
Green's function
\begin{equation}
G_{ij}(t)=-i\theta(t)\langle \{c_i(t),c_j^{\dagger}(0)\}\rangle
\label{gret}
\end{equation}
and its Fourier transform
\begin{equation}
G_{ij}(E)=\int_{-\infty}^{\infty} dt \, G_{ij}(t) e^{iEt}.
\label{gret.ft}
\end{equation}
The Green's function of the tunneling Hamiltonian $H_0$ satisfies
\begin{equation}
G^{-1}_0(E) = {\bf 1} (E+i0^+) - H_0,
\label{g0}
\end{equation}
where ${\bf 1}$ is the unit matrix.
In the $|S\rangle,\,|N\rangle$ basis, one has
\begin{equation}
G_0^{-1}(E) 
= \left(\begin{array}{cc}
E-\varepsilon_S+i0^+& -V \\ -V & E-\varepsilon_N+i0^+\end{array}\right).
\label{g0.2}
\end{equation}
The full Green's function, including decay processes, may be calculated from
Dyson's equation
\begin{equation}
G^{-1} = G_0^{-1} - \Sigma,
\end{equation}
where $\Sigma$ is the self-energy matrix describing the decay processes
$\Gamma_S$ and $\Gamma_N$ induced by $H_D$.  
The simplest ansatz for $\Sigma$ is \cite{wbb}
\begin{equation}
\Sigma \equiv \left(\begin{array}{cc} \Sigma_{SS} & \Sigma_{SN} \\
\Sigma_{NS} & \Sigma_{NN} \end{array}\right)
= \left(\begin{array}{cc}
-i\Gamma_S/2 & 0 \\ 0 & -i\Gamma_N/2 \end{array}\right).
\label{sigma}
\end{equation}
Using Eqs.\ (\ref{g0.2}) and (\ref{sigma}), one can solve
Dyson's equation 
to obtain the full retarded Green's function of the two-level system,
\begin{eqnarray}
G  & \equiv & \left(\begin{array}{cc} G_{SS} & G_{SN} \\
G_{NS} & G_{NN} \end{array}\right)
\nonumber \\
&  = & [(E-\varepsilon_S+i\Gamma_S/2)(E-\varepsilon_N+i\Gamma_N/2)-V^2]^{-1}
\nonumber \\
&& \times \left(\begin{array}{cc} E-\varepsilon_N+i\Gamma_N/2 & -V \\
-V & E-\varepsilon_S+ i\Gamma_S/2\end{array}\right).
\label{gfull}
\end{eqnarray}
$\mbox{}$From $G$, all information about the dynamics of the system
and the branching ratios of the decay processes can be obtained.

Let us first study the dynamics of the coupled SD--ND system.
Assuming the nucleus starts out at time zero in the superdeformed
state $|S\rangle$, 
the probability that the nucleus is in state $|S\rangle$ at a later
time $t$ is
$P_S(t)=|G_{SS}(t)|^2$.  The probability that the nucleus is in the normal
state $|N\rangle$ at time $t$ is $P_N(t)=|G_{NS}(t)|^2$.  
$P_S(t)$ and $P_N(t)$ may be calculated straightforwardly from the 
Fourier transform of Eq.\ (\ref{gfull}).  The general result for 
$P_N(t)$ is
\begin{equation}
P_N(t)= \frac{2V^2}{|\omega|^2} 
e^{-\bar{\Gamma}t} \left(\cosh \omega_i t - \cos \omega_r t\right),
\label{P_N.gen}
\end{equation}
where
\begin{equation}
\omega \equiv \omega_r + i \omega_i = \sqrt{4V^2 + (\Delta - i\Gamma')^2},
\label{omega}
\end{equation}
with $\Delta=\varepsilon_N-\varepsilon_S$
and $\Gamma'=(\Gamma_N-\Gamma_S)/2$.  The general expression for $P_S(t)$
is rather lengthy.

The tunneling dynamics is particularly interesting when the energy difference
$\Delta=0$.
There are then two qualitatively different dynamical regimes, depending on the 
relative size of the tunneling matrix element $V$ and the difference
in decay rates $\Gamma'$.
For $2V > |\Gamma'|$, the tunneling dynamics is {\em coherent}, and
one finds
\begin{eqnarray}
P_S(t) & = & 
e^{-\bar{\Gamma}t} \left[\cos^2\frac{\omega_0 t}{2}
+ \frac{\Gamma'}{\omega_0}\sin\omega_0 t +
\frac{\Gamma'^2}{\omega_0^2}\sin^2 \frac{\omega_0 t}{2}\right],
\nonumber\\
P_N(t) & = & 
\frac{4V^2}{\omega_0^2} 
e^{-\bar{\Gamma}t}
\sin^2 \frac{\omega_0 t}{2},
\label{P.coh}
\end{eqnarray}
where the Rabi frequency $\omega_0$ is 
\begin{equation}
\omega_0=\left|4V^2- \Gamma'^2\right|^{1/2}.
\label{freq.rabi}
\end{equation}
For $2V < |\Gamma'|$, on the other hand,
the tunneling dynamics is {\em incoherent}, and
\begin{eqnarray}
P_S(t) & = & 
e^{-\bar{\Gamma}t} \left[\cosh^2\frac{\omega_0 t}{2}
+ \frac{\Gamma'}{\omega_0}\sinh\omega_0 t +
\frac{\Gamma'^2}{\omega_0^2}\sinh^2 \frac{\omega_0 t}{2}\right],
\nonumber\\
P_N(t) & = & 
\frac{4V^2}{\omega_0^2} 
e^{-\bar{\Gamma}t}
\sinh^2 \frac{\omega_0 t}{2}.
\label{P.incoh}
\end{eqnarray}
For $\Gamma'=0$ and $\Delta \neq 0$, the tunneling dynamics
is {\em coherent},
given by Eq.\ (\ref{P.coh}) with 
$\omega_0 \rightarrow (4V^2+\Delta^2)^{1/2}$.
For $\Gamma'\neq 0$ and $\Delta \neq 0$, 
the tunneling dynamics has
both coherent and incoherent components 
[c.f.\ Eq.\ (\ref{P_N.gen})], the coherent component being suppressed for
large $\Gamma'$ and/or large $\Delta$.

The dynamics of the system is similar to that of the two-level system
with dissipation, investigated by Leggett and collaborators \cite{leggett};  
the principal difference is that we consider a two-level system in which the 
total number of particles is itself time-dependent.
The physical origin of the imaginary 
self energy $\Sigma$ is virtual transitions of the nucleus
to lower-lying states and back again, 
which alter the state of the electromagnetic environment.
This is analogous to the coupling of the two-level system to a bath of  bosonic
excitations considered in Ref.\ \onlinecite{leggett}.  If the environment
couples with equal strength to the states $S$ and $N$, i.e., if $\Gamma_S=
\Gamma_N=\bar{\Gamma}$, the Green's function factorizes quite generally
\cite{casnsw}:  $G(t) = e^{-\bar{\Gamma}t/2}G_0(t)$, and the nucleus 
undergoes Rabi oscillations with frequency $\omega=(4V^2+\Delta^2)^{1/2}$ 
between the states $S$
and $N$.  The nucleus is in a coherent superposition of states, which decays
at an overall rate $\bar{\Gamma}$ to lower-lying states.  
However, if the environment
couples with different strengths to the states $S$ and $N$, i.e., if  
$\Gamma'\neq 0$, coherent tunneling between $S$ and $N$ is 
suppressed since the environment ``measures'' which state the system is in.  
For $\Delta=0$ and $0<|\Gamma'|<2V$, 
the dynamics described by Eq.\ (\ref{P.coh})
is qualitatively similar to the case $\Gamma'=0$, but the Rabi frequency
is reduced to the value given in Eq.\ (\ref{freq.rabi}).  
If the difference in coupling exceeds the critical value
$|\Gamma'|> 2V$ for $\Delta=0$, the coherent superposition of $S$ and $N$ is
destroyed altogether, and the dynamics is overdamped.  
As in the model of Ref.\ \onlinecite{leggett}, there are both coherent and
incoherent components of the dynamics when both
$\Gamma'\neq 0$ and $\Delta\neq 0$.

Let us next turn our attention to the decay branching ratio, which is the
experimentally measurable quantity. 
When the nucleus is in the 
state $S$, it decays at a rate $\Gamma_S$ to a lower superdeformed state, and
when it is in state $N$ it decays at a rate $\Gamma_N$ to a lower-energy
state in the normal-deformed band.  Thus, the time-dependent rates to decay in 
the $S$ and $N$ channels are 
\begin{equation}
\tilde{\Gamma}_S(t)=\Gamma_S P_S(t), 
\;\;\;
\tilde{\Gamma}_N(t)=\Gamma_N P_N(t). 
\label{eq.decay}
\end{equation}
The fraction $F_N$
of nuclei that decay via E1 processes in the normal-deformed band
is just \cite{parseval}
\begin{equation}
F_N = \frac{\int_0^{\infty} dt \, \tilde{\Gamma}_N(t)}{\int_0^{\infty} dt \, 
[\tilde{\Gamma}_N(t)+\tilde{\Gamma}_S(t)]}
=\Gamma_N\int_0^{\infty} dt\, P_N(t).
\label{fN1}
\end{equation}
This integral may be evaluated 
to obtain the central result of this paper,
\begin{equation}
F_N = \frac{(1+\Gamma_N/\Gamma_S)V^2}{\Delta^2 + \bar{\Gamma}^2
(1+4V^2/\Gamma_N\Gamma_S)}.
\label{fN.result}
\end{equation}
The fraction of nuclei decaying via E2 processes within the superdeformed band 
is $F_S=1-F_N$.  In Ref.\ \onlinecite{wbb}, $F_S$ was denoted by $F$.

Let us next consider some limits of Eq.\ (\ref{fN.result}).  In the limit of
very strong coupling of the states $S$ and $N$, $V \gg \bar{\Gamma}$, 
one finds
\begin{equation}
\lim_{V/\bar{\Gamma}\rightarrow \infty} F_N =
\frac{\Gamma_N(\Gamma_S+\Gamma_N)}{(\Gamma_S+\Gamma_N)^2+\Gamma_S\Gamma_N
(\Delta/V)^2}.
\label{fS.large}
\end{equation}
This is equivalent to the result of Vigezzi, Broglia, and Dossing
\cite{vigezzi1,vigezzi2} 
for the case
where only a single SD and ND state mix:  
\begin{equation}
F_N^{\rm vbd} = \sum_{\sigma=\pm} \frac{|c_\sigma|^2
(1-|c_\sigma|^2) \Gamma_N}{
(1-|c_\sigma|^2)\Gamma_N + |c_\sigma|^2 \Gamma_S},
\label{fN.vig}
\end{equation}
where $c_\pm = \langle \pm|S\rangle$ are the mixing amplitudes of the
eigenstates $|\pm\rangle$ of $H_0$ with $|S\rangle$.  From Eq.\ (\ref{h0}),
we have
\begin{equation}
|c_\pm|^2 = [1+(x \pm \sqrt{x^2+1})^2]^{-1},
\label{amplitude}
\end{equation}
with $x=\Delta/2V$.  Inserting Eq.\ (\ref{amplitude}) into Eq.\ (\ref{fN.vig}),
one indeed recovers Eq.\ (\ref{fS.large}).
Thus the result (\ref{fN.vig}) of Refs.\ \onlinecite{vigezzi1,vigezzi2}
is seen to be a limiting case for $V/\bar{\Gamma}\rightarrow \infty$ 
(i.e., for {\em fully coherent} tunneling) of
our general result, Eq.\ (\ref{fN.result}).

Another limit was considered in Ref.\ \onlinecite{wbb}, namely
$\Gamma_N \gg \Gamma_S,\,\Gamma^\downarrow$.
In this limit, the tunneling dynamics is {\em incoherent}.
The assumption $\Gamma_S \ll \Gamma_N$ is motivated by the fact that
$\Gamma_S$ is an E2 decay and $\Gamma_N$ is an E1 decay.  
In the limit $\Gamma_N\gg\Gamma_S$,
Eq.\ (\ref{fN.result}) simplifies to
\begin{equation}
\lim_{\Gamma_N/\Gamma_S\rightarrow \infty}
F_N = \frac{\Gamma^{\downarrow}}{\Gamma_S+\Gamma^{\downarrow}},
\label{fS.small}
\end{equation}
where $\Gamma^{\downarrow}$ is given by Eq.\ (\ref{gamma.right}).  Eq.\
(\ref{fS.small}) is 
identical to the principal result of Weidenm\"uller, von Brentano, and
Barrett \cite{wbb}, although our expression for $\Gamma^\downarrow$ differs
from that of Ref.\ \onlinecite{wbb}.  
Note that, in contrast to the argument of Ref.\
\onlinecite{wbb}, no assumption has been made about the relative size
of $V$ and $\bar{\Gamma}$ in deriving Eq.\ (\ref{fS.small}) from
Eq.\ (\ref{fN.result}).  However, the interpretation of $\Gamma^\downarrow$
as a {\em tunneling rate} is only justified when $\Gamma^\downarrow/
\bar{\Gamma}\ll 1$; for larger values of $V$, the tunneling dynamics
described by Eq.\ (\ref{P_N.gen}) is more complex, though the integrated 
rate still obeys Eq.\ (\ref{fS.small}), provided $\Gamma_N\gg \Gamma_S$.

Eq.\ (\ref{gamma.right}) for $\Gamma^{\downarrow}$
is also the expression one would obtain from a 
correct application of Fermi's golden rule in the limit $V \ll \bar{\Gamma}
\ll D_N$: 
\begin{equation}
\Gamma^\downarrow = 2\pi V^2 \int_{-\infty}^\infty
dE \rho_S(E) \rho_N(E),
\label{fgr2}
\end{equation}
where the lifetime-broadened densities-of-states of the SD and ND levels
are
\begin{eqnarray*}
\rho_S(E) & = & \frac{\Gamma_S/2\pi}{(E-\varepsilon_S)^2+\Gamma_S^2/4},
\label{rho_s}\\
\nonumber\\
\rho_N(E) & = & \frac{\Gamma_N/2\pi}{(E-\varepsilon_N)^2+\Gamma_N^2/4}.
\label{rho_n}
\end{eqnarray*}
Evaluating the integral in Eq.\ (\ref{fgr2}), the 
expression (\ref{gamma.right}) is obtained.
The level-spacing 
$D_N$ in the ND band is irrelevant if $V\ll D_N$, since $V$ only 
mixes the state $S$ and  the single state $N$ which is closest to it in 
energy in that case, as we have assumed.

The branching ratio (\ref{fN.result}) depends strongly on the energy 
difference $\Delta =\varepsilon_N-\varepsilon_S$, which in turn depends
sensitively on the microscopic Hamiltonian of the particular nucleus under
investigation.  In order to eliminate this parameter dependence, one 
practice which is employed is to calculate the average of $F_N$ over
a statistical (GOE) ensemble of similar nuclei.   
In the limit of incoherent tunneling $V\ll\Gamma_S,\,\Gamma_N$,
Eq.\ (\ref{fN.result}) may be integrated over $\Delta$ to obtain
\begin{equation}
\langle F_N \rangle = \int_{-\infty}^\infty \frac{d\Delta}{D_N} F_N(\Delta)
=
\frac{\langle \Gamma^\downarrow\rangle}{\Gamma_S},
\label{fN.ave}
\end{equation}
where $\langle \Gamma^\downarrow\rangle$ is given by the right-hand-side of
Eq.\ (\ref{gamma.wrong}), in agreement with Ref.\ \onlinecite{wbb}. 
However, it is clear from Eq.\ (\ref{fN.result}) that $F_N$ typically 
deviates significantly from its mean value.  For instance, the mean
square of $F_N$ is much larger than the square of $\langle F_N\rangle$
when $D_N \gg \bar{\Gamma}$:
\begin{equation}
\frac{\langle F_N^2\rangle}{\langle F_N\rangle^2} = 
\frac{D_N \langle V^4\rangle}{2\pi \bar{\Gamma} \langle V^2\rangle^2}.
\label{fN.sqave}
\end{equation}
Thus, it would be preferable to compare experimental results directly with
Eq.\ (\ref{fN.result}), rather than with its ensemble average.

In Table I, we show some experimental data for nuclei in the 
$A\sim 150$ and $A\sim 190$ mass regions.  So far, little data is available
in the $A\sim 150$ region, with only estimates \cite{vigezzi1,krucken}
for the widths for $^{152}$Dy.  From the numbers given in Table I, we 
observe that the nuclei in the $A\sim 190$ mass region have
$\Gamma^{\downarrow},\, \Gamma_S\ll\Gamma_N$.  The dynamics of $S\rightarrow
N$ tunneling
in these nuclei is thus incoherent, and the appropriate branching ratio
is given by Eq.\ (\ref{fS.small}), in agreement with Ref.\ \onlinecite{wbb}.
On the other hand, for $^{152}$Dy, $\Gamma^{\downarrow} \gg \Gamma_S,\Gamma_N$,
indicating  coherent $S\rightleftharpoons N$ 
tunneling.  The measured value \cite{krucken} of 
$F_S=0.51$ for $^{152}$Dy(26) is
consistent with Eq.\ (\ref{fS.large}), using the values of $\Gamma_S$ and
$\Gamma_N$ in Table I and assuming $V/\Delta\sim 1$, 
in accord with the theory of Refs.\ 
\onlinecite{schiffer,vigezzi1,vigezzi2,shimizu,shimizu2,khoo}.  Our general
result (\ref{fN.result}) is consistent with the data in both the 
$A\sim 150$ and $A\sim 190$ mass regions and unifies these two complementary 
theoretical approaches.

A recent paper \cite{aberg} shows that the 
$S\rightarrow N$
tunneling rate may be enhanced by several orders of magnitude if the 
ND states are chaotic at the moment of the decay out.  These results 
are not inconsistent with our two-level model, but would simply imply an
enhancement of the tunneling matrix element $V$.

In conclusion, we have shown that a simple two-level model, which 
treats decay and tunneling processes on an equal footing, can explain the 
apparently disparate previous theoretical results, i.e., Refs.\
\onlinecite{schiffer,vigezzi1,vigezzi2,shimizu,shimizu2,khoo} versus
Ref.\ \onlinecite{wbb}, for the decay out of a superdeformed band.  These
previous results are shown to correspond to the coherent and incoherent
limits, respectively, of the tunneling dynamics, and are special 
cases of our general result, Eq.\ (\ref{fN.result}).
We remark that
it is straightforward to extend our method to treat an SD state coupled to
an arbitrary number of ND states.

BRB thanks R. Kr\"ucken for helpful discussions regarding the
experimental data for $^{152}$Dy, and acknowledges partial support
from NSF grant No.\ PHY9605192.

\begin{table}[b]
\caption{Widths and level spacings for a number of nuclei, deduced from
the data of Refs.\ [2,8], following the 
analysis of Refs.\ [2,12] for $A\sim 150$ and of
Ref.\ [7] for $A\sim 190$.  The spin  values
of the decaying states are given in parentheses.}
\label{table:widths}
\begin{tabular}{ccccc}
Nucleus & $\Gamma_N$ (meV)  & $D_N$ (eV) & 
$\Gamma_S$ (meV) & $\Gamma^{\downarrow}$ (meV)
\\
\tableline
$^{152}$Dy(26) & 10--20 & 3--10 & $\sim 11$ & 900--3000\\
$^{152}$Dy(24) & 10--20 & 3--10 & $\sim 7.6$ & 900--3000\\
$^{192}$Hg(12) & 10.3 & 34 & 0.116 & 0.018\\
$^{192}$Hg(10) & 10.3 & 30 & 0.054 & 0.544\\
$^{194}$Hg(12) & 18.1 & 92 & 0.144 & 0.097\\
$^{194}$Hg(10) & 18.4 & 79 & $\geq 0.047$ & $\geq 0.89$\\
$^{194}$Pb(10) & 1.6 & 1699 & 0.066 & 0.011\\
$^{194}$Pb(8) & 1.7 & 1549 & 0.028 & 0.009\\
\end{tabular}
\end{table}


\begin{references}
\bibitem{parseval} It should be mentioned that the branching ratios can
also be expressed directly in terms of the energy Green's functions
using Parseval's theorem:
\[
F_N\!=\Gamma_N\!\int_{-\infty}^{\infty} \frac{dE}{2\pi} |G_{NS}(E)|^2\!, \;
F_S\!=\Gamma_S\!\int_{-\infty}^{\infty} \frac{dE}{2\pi} |G_{SS}(E)|^2.
\]
\end{references}
\end{document}